\documentclass[3p,times,procedia]{elsarticle}
\usepackage{nupha_ecrc}

\volume{00}

\firstpage{1}

\journalname{Nuclear Physics A}

\runauth{M. Varga-Kofarago}

\jid{nupha}

\jnltitlelogo{Nuclear Physics A}

\usepackage{amssymb}

\usepackage[figuresright]{rotating}

\usepackage{graphicx}
\usepackage{overpic}
\usepackage{subfig}
\usepackage{amsmath}
\usepackage[capitalise]{cleveref}

\def\Plus{\texttt{+}}
\def\Minus{\texttt{-}}

\begin{document}

\begin{frontmatter}



\dochead{XXVIIth International Conference on Ultrarelativistic Nucleus-Nucleus Collisions\\ (Quark Matter 2018)}

\title{The evolution of the near-side peak in two-particle number and transverse momentum correlations in Pb--Pb collisions from ALICE}


\author{Monika Varga-Kofarago \\  \normalsize (varga-kofarago.monika@wigner.mta.hu) \\ \vspace{1mm} \normalsize{on behalf of the ALICE collaboration}}

\address{Hungarian Academy of Science Wigner RCP}

\begin{abstract}
Two-particle number and transverse momentum correlations are powerful tools for studying the medium produced in heavy-ion collisions. Correlations in the angular separation of pairs of hadrons can provide information on the medium transport characteristics.  In particular, the transverse momentum correlations are sensitive to momentum currents, and provide information about the system life time, the shear viscosity over entropy density ratio ($\eta/s$) and the system relaxation time ($\tau_{\pi}$). Furthermore, the interaction of the jets produced in the initial stages of a collision can be studied using number correlations, by observing the medium-induced modification of the near-side jet peak. Measurements of both sets of correlations from Pb--Pb collisions are reported as a function of centrality. Theoretical interpretations and results from Monte Carlo generators are then confronted with the experimental data.
\end{abstract}


\end{frontmatter}


\section{Introduction}
The medium produced in heavy-ion collisions, the Quark--Gluon Plasma (QGP), can be studied by two-particle angular correlation measurements. In these measurements, the azimuthal angle ($\Delta\varphi$) and pseudorapidity ($\Delta\eta$) differences of two particles are calculated and different properties of the pair are studied as a function of these variables. On the one hand, looking at the centrality and momentum dependence of two-particle number correlations, one can study the interactions of jets with the QGP in a momentum regime where direct jet reconstruction is not possible because of the large fluctuating background of heavy-ion collisions~\cite{PRL,PRC}. Transverse momentum correlations, on the other hand, are sensitive to the dynamics of the collision, and the centrality evolution of the near-side peak in these correlations carries information on the shear viscosity over entropy density ratio ($\eta/s$) and the relaxation time ($\tau_\pi$) of the system~\cite{STAR}. 

In two-particle number correlations, the per-trigger yield is calculated as
\begin{equation}
  \frac{1}{N_{\rm trig}}\frac{\text{d}^2N_{\rm assoc}}{\text{d}\Delta\varphi \text{ d}\Delta\eta } = \frac{S(\Delta\varphi,\Delta\eta)}{M(\Delta\varphi,\Delta\eta)},
\end{equation}
where $N_{\rm trig}$ is the number of trigger particles, $S(\Delta\varphi,\Delta\eta)$ stands for the yield of particle pairs in the case the particles are taken from the same event, while $M(\Delta\varphi,\Delta\eta)$ is referred to as the mixed event and represents the same yield constructed such that the particles are taken from different events. By this division the effects of the limited acceptance of the detector can be corrected for. In transverse momentum ($p_{\rm T}$) correlations, the following quantity is calculated
\begin{equation}
  G_2(\Delta\varphi,\Delta\eta) = \frac{\left<\sum\limits_{i}^{n_{1,1}}\sum\limits_{j \neq i}^{n_{1,2}}p_{\rm T,i}p_{\rm T,j}\right>}{\left<n_{1,1}\right>\left<n_{1,2}\right>}-\left<p_{\rm T,1}\right>\left<p_{\rm T,2}\right>.
\end{equation}
In this equation, $p_{\rm T}$ is the transverse momentum of the particles and $n_1$ is the number of particles at a given $\Delta\varphi$ and $\Delta\eta$. The summation goes over all particles of the event and the angled brackets symbolize the event averages. This correlator is constructed such that if there is no correlation between the momenta of the particles, it gives zero. In two-particle number correlations, all charged hadrons are considered when the per-trigger yield is calculated, while in the case of transverse momentum correlations, like-sign and unlike-sign pairs are treated separately. From these, the so-called charge-dependent (CD: $\frac{1}{4}~[(+-)~\Plus~(-+)~\Minus~(++)~\Minus~(--)]$) and charge-independent cases (CI: $\frac{1}{4}~[(+-)~\Plus~(-+)~\Plus~(++)~\Plus~(--)]$) are built.

\begin{figure}[!htbp]
  \makebox[\textwidth][c]{\hfill
    \begin{overpic}[width=0.35\textwidth]{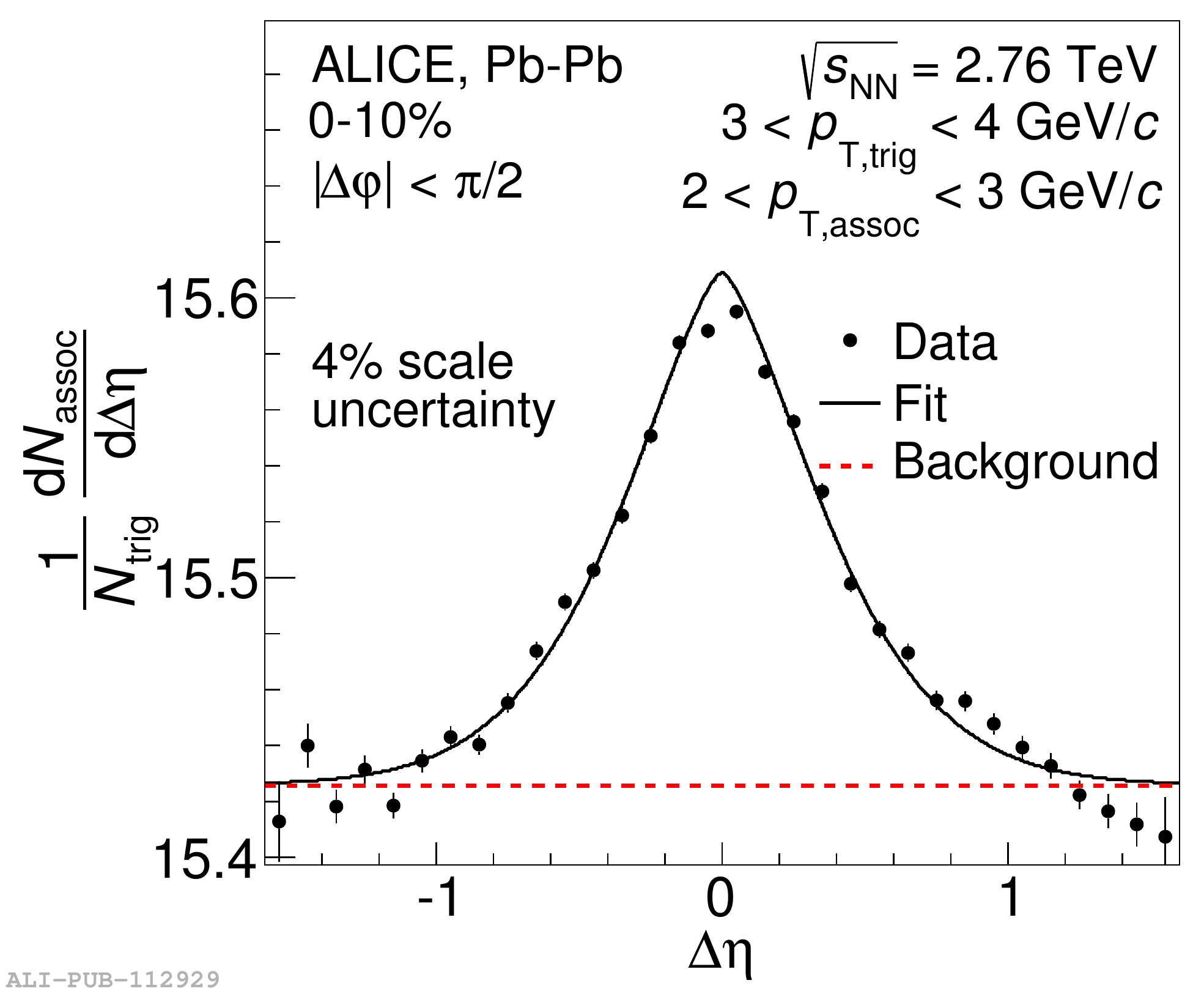}
    \end{overpic}\hfill
    \begin{overpic}[width=0.4\textwidth]{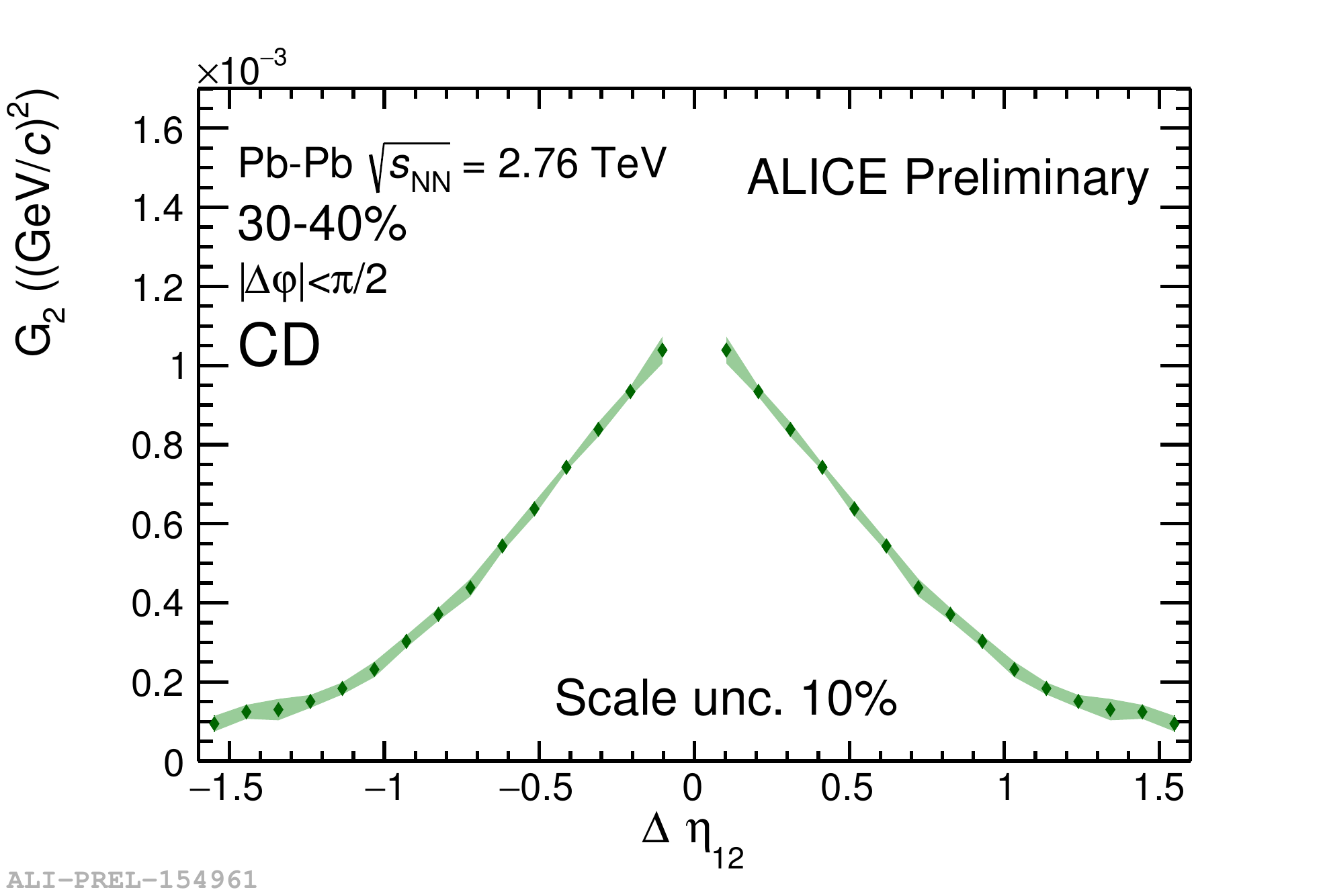}
    \end{overpic}\hfill}
  \caption{The left panel shows the projection to the $\Delta\eta$ axis of a chosen centrality and $p_{\rm T}$ bin of number correlations, while the right panel shows a chosen centrality bin of the charge-dependent transverse momentum correlations from Pb--Pb collisions at $\sqrt{s_ {\rm NN}} = 2.76$~TeV. The green band in the right panel represents the point-by-point systematic uncertainties.}
  \label{fig:proj}
\end{figure}

In \cref{fig:proj}, the projections to the $\Delta\eta$ axis of a chosen $p_{\rm T}$ and centrality bin of number correlations and of a chosen centrality bin of the charge-dependent transverse momentum correlations are shown. Both show a peak with similar shape around $\Delta\eta = 0$ and a constant background. In the $\Delta\varphi$ direction, both correlations show a peak around $\Delta\varphi = 0$, which is sitting on top of the background modulated by anisotropic flow. As a consequence of this similarity in the shape of the correlations, both of them can be fitted in two dimensions by the sum of a generalized Gaussian function, a constant and cosine functions to account for the flow part
\begin{equation}
  C_1 + \sum\limits_{n=2}^N 2 V_n \cos (n \Delta\varphi) + C_2\times \frac{\gamma_\varphi\gamma_\eta}{4w_{\varphi}w_{\eta}\Gamma\left(\frac{1}{\gamma_\varphi}\right)\Gamma\left(\frac{1}{\gamma_\eta}\right)}\times e^{-\left|\frac{d\varphi}{w_\varphi}\right|^{\gamma_\varphi}-\left|\frac{d\eta}{w_\eta}\right|^{\gamma_\eta}}.
\end{equation}
From these, the variance of the generalized Gaussian is extracted and in the following this will be quoted as the width of the near-side peak.

\section{Results}

\begin{figure}[!htbp]
  \makebox[\textwidth][c]{\hfill
    \begin{overpic}[width=0.45\textwidth]{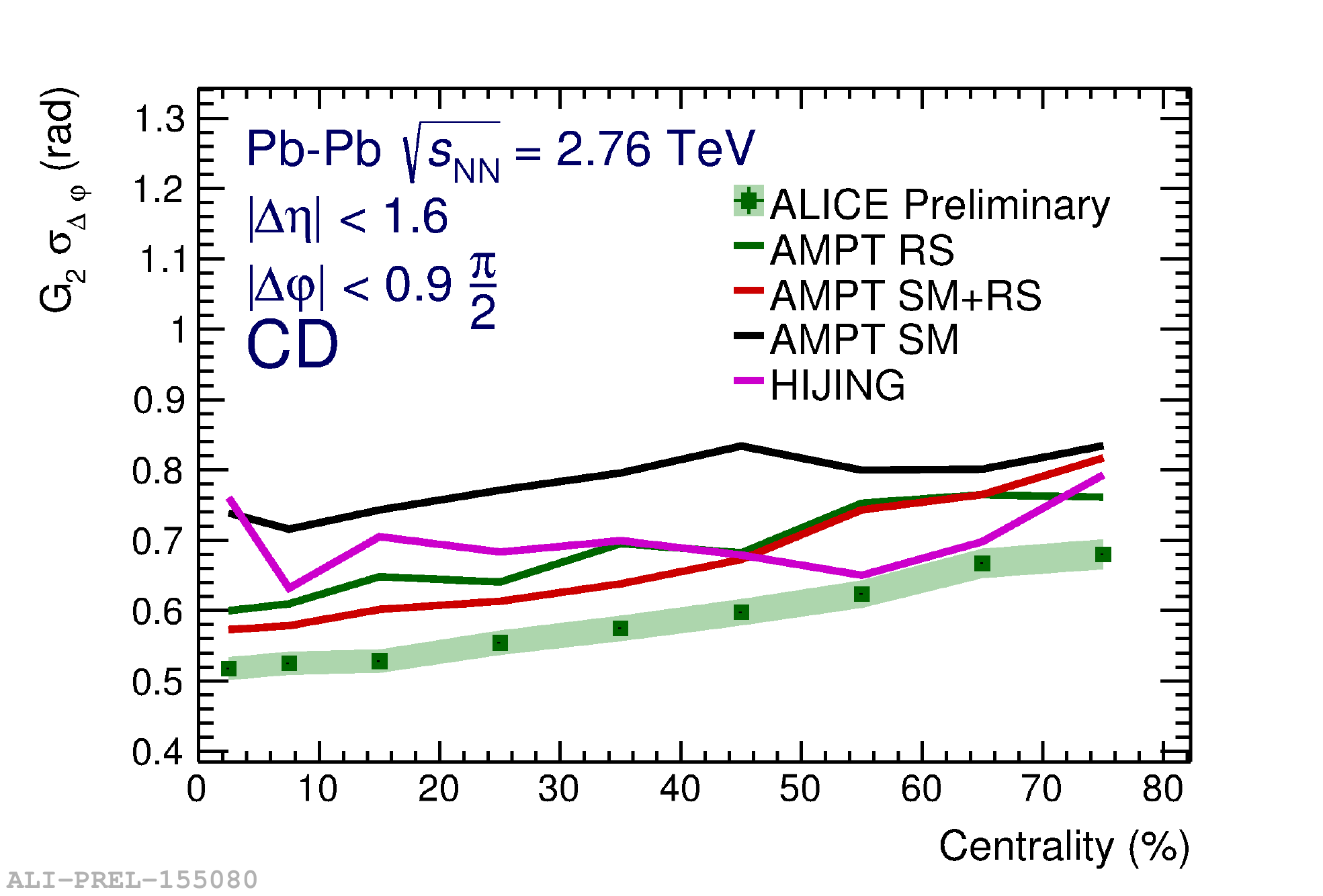}
    \end{overpic}
    \begin{overpic}[width=0.45\textwidth]{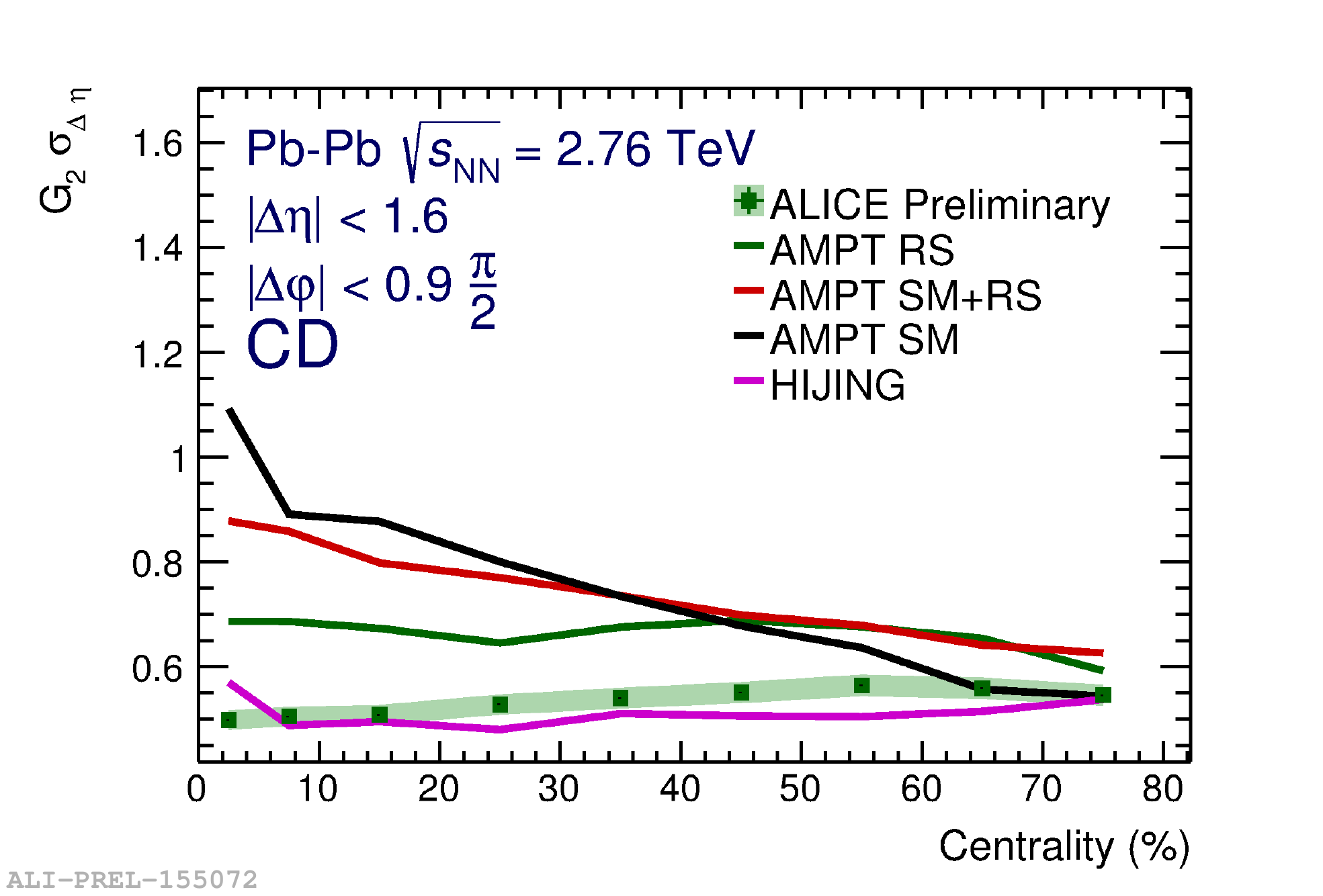}
    \end{overpic}}

  \makebox[\textwidth][c]{\hfill
    \begin{overpic}[width=0.45\textwidth]{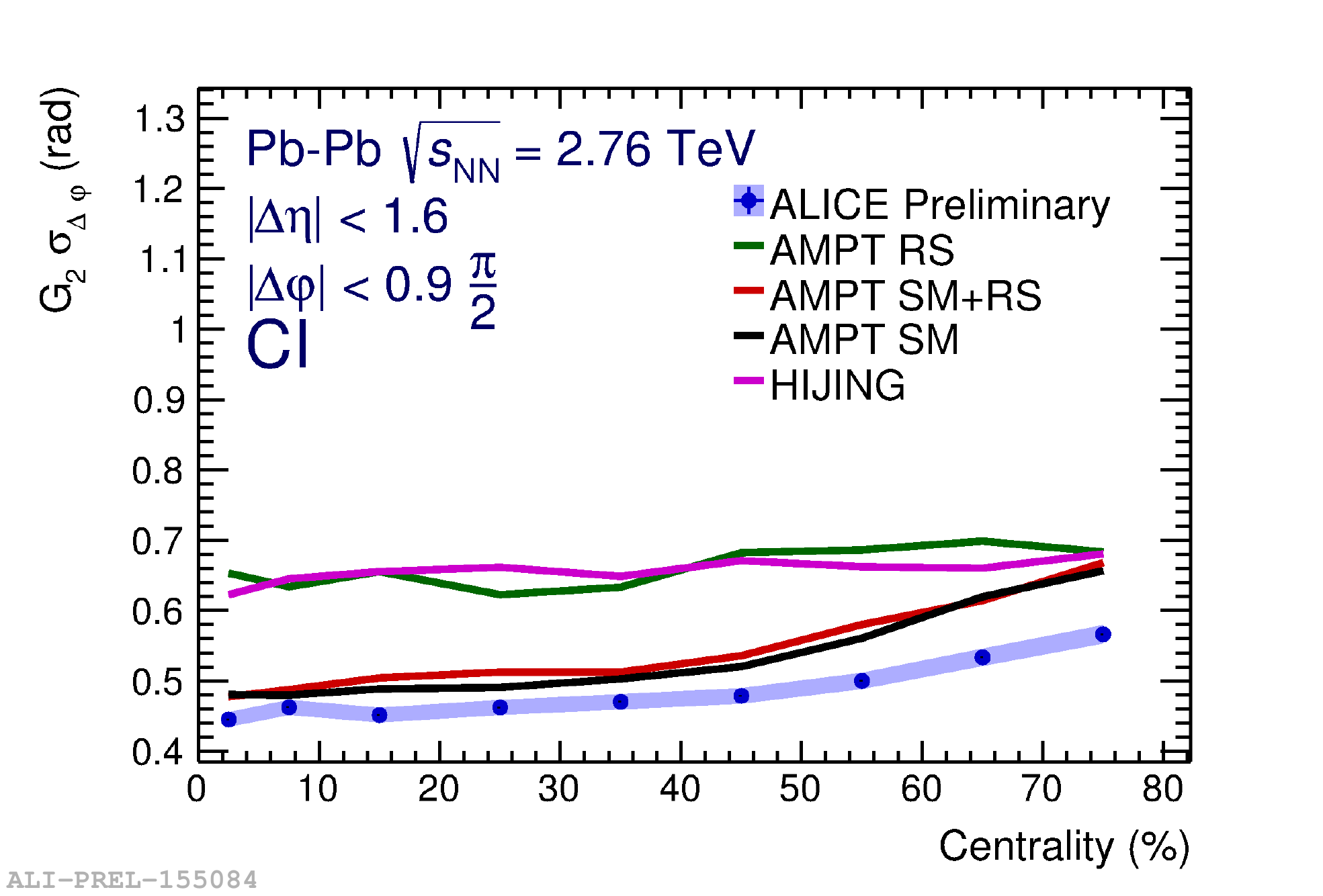}
    \end{overpic}
    \begin{overpic}[width=0.45\textwidth]{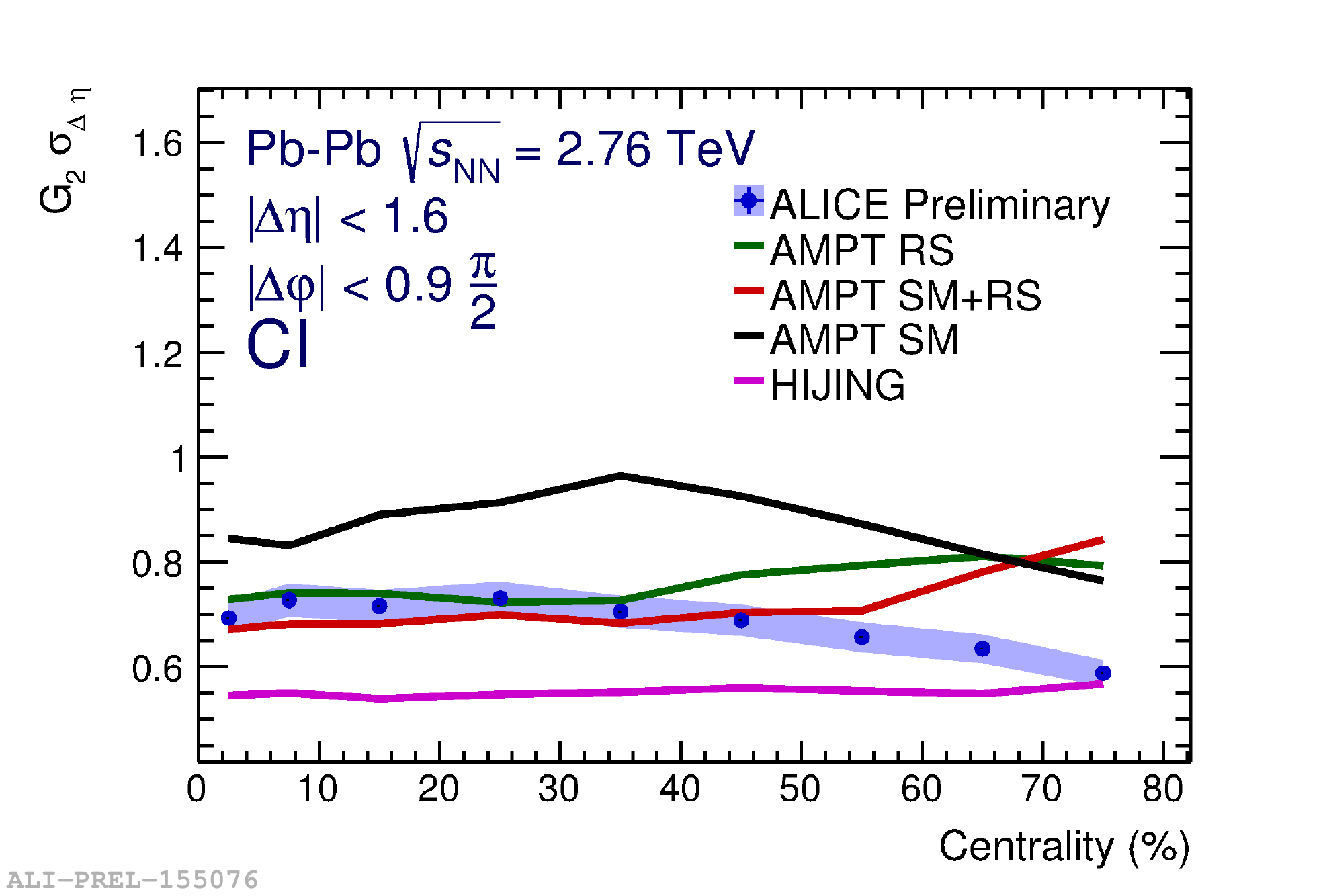}
    \end{overpic}}
  \caption{The width of the near-side peak in transverse momentum correlations as a function of centrality in the $\Delta\varphi$ direction (left panels) and the $\Delta\eta$ direction (right panels). The top panels show the charge-dependent case, while the bottom panels show the charge-independent case. Each panel shows the data from $\sqrt{s_ {\rm NN}} = 2.76$~TeV Pb--Pb collisions as markers with the green and blue bands representing the systematic uncertainties. Results from HIJING and AMPT simulations are overlaid with the data, and the different settings of AMPT are explained in \cref{sec:MC}.}
  \label{fig:width_pT}
\end{figure}

In \cref{fig:width_pT}, the widths of the near-side peak in transverse momentum correlations are shown. Both the charge-dependent and charge-independent near-side peak narrows azimuthally with the centrality of the collision (left panels), while longitudinally (right panels), the charge-independent broadens with collision centrality and the charge-dependent stays almost unaffected. The broadening with centrality of the charge-independent transverse momentum correlation was connected to the shear viscosity over entropy density ratio of the system by the STAR collaboration~\cite{STAR}.

\begin{figure}[!htbp]
  \makebox[\textwidth][c]{
    \begin{overpic}[width=0.55\textwidth]{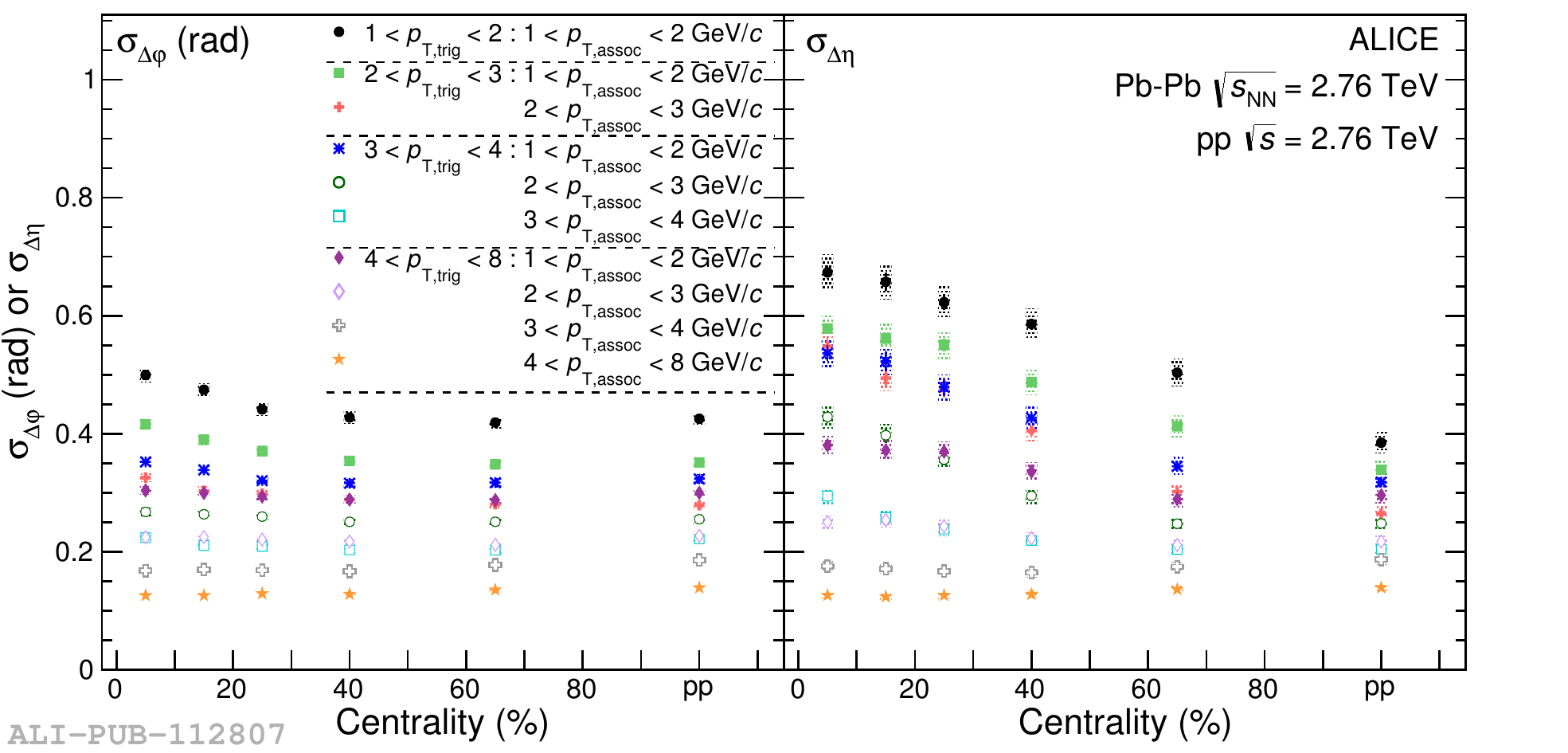}
    \end{overpic}
    \begin{overpic}[width=0.55\textwidth]{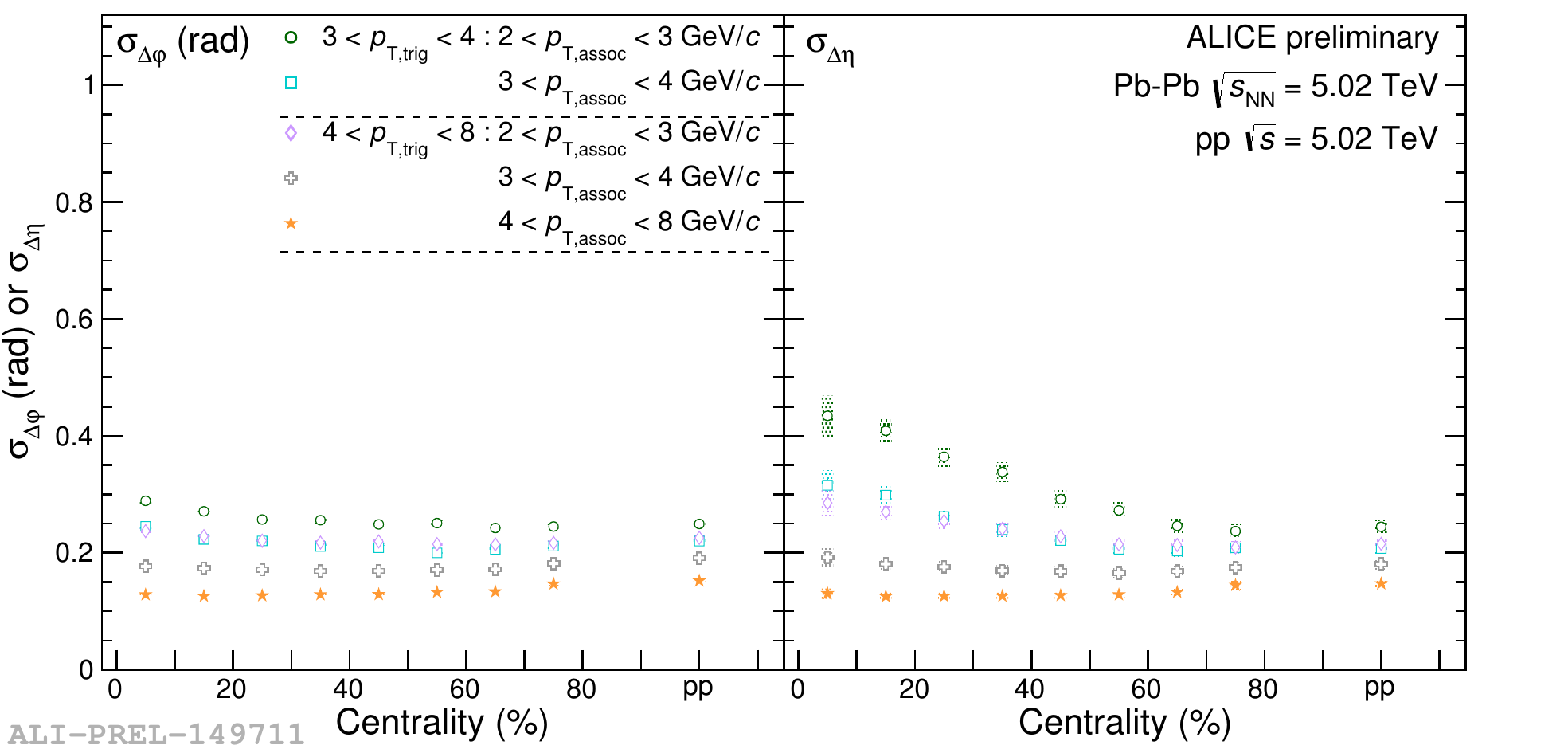}
    \end{overpic}}
  \caption{The width of the near-side peak in two-particle number correlations as a function of centrality from Pb--Pb collisions with the right-most point in each panel showing the results from pp collisions. The left figure shows the results from $\sqrt{s_ {\rm NN}} = 2.76$~TeV, while the right figure shows the results from $\sqrt{s_ {\rm NN}} = 5.02$~TeV collisions.}
  \label{fig:width_number}
\end{figure}

In \cref{fig:width_number}, the width of the near-side peak in number correlations is shown. The width shows a moderate broadening at low $p_{\rm T}$ towards central events in the $\Delta\varphi$ direction and a much more pronounced broadening in the $\Delta\eta$ direction. This broadening is present at both energies ($\sqrt{s_ {\rm NN}} = 2.76$~TeV and $\sqrt{s_ {\rm NN}} = 5.02$~TeV).

\begin{figure}[!htbp]
  \makebox[\textwidth][c]{\hfill
    \begin{overpic}[width=0.35\textwidth]{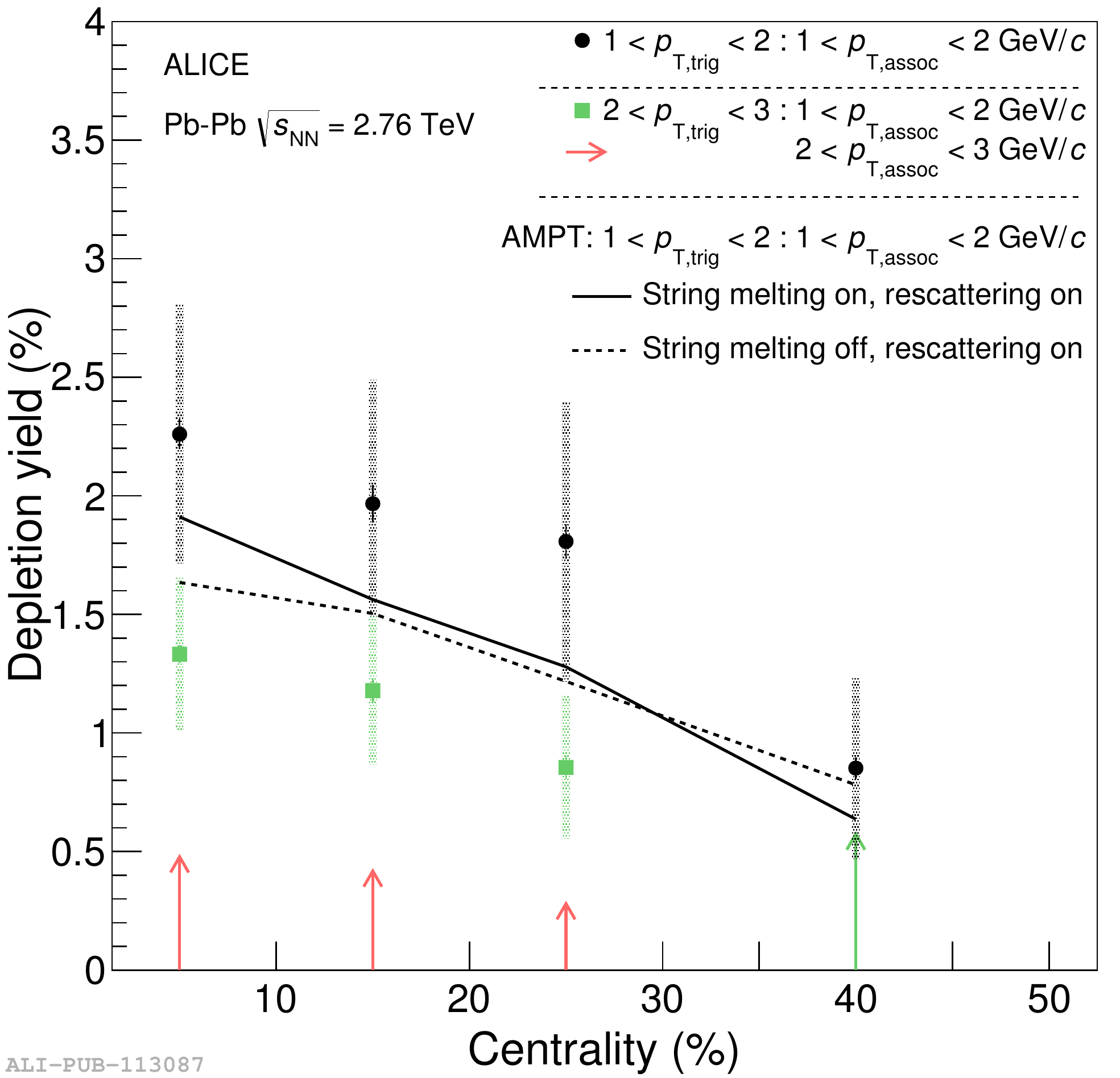}
    \end{overpic}\hfill
    \begin{overpic}[width=0.35\textwidth]{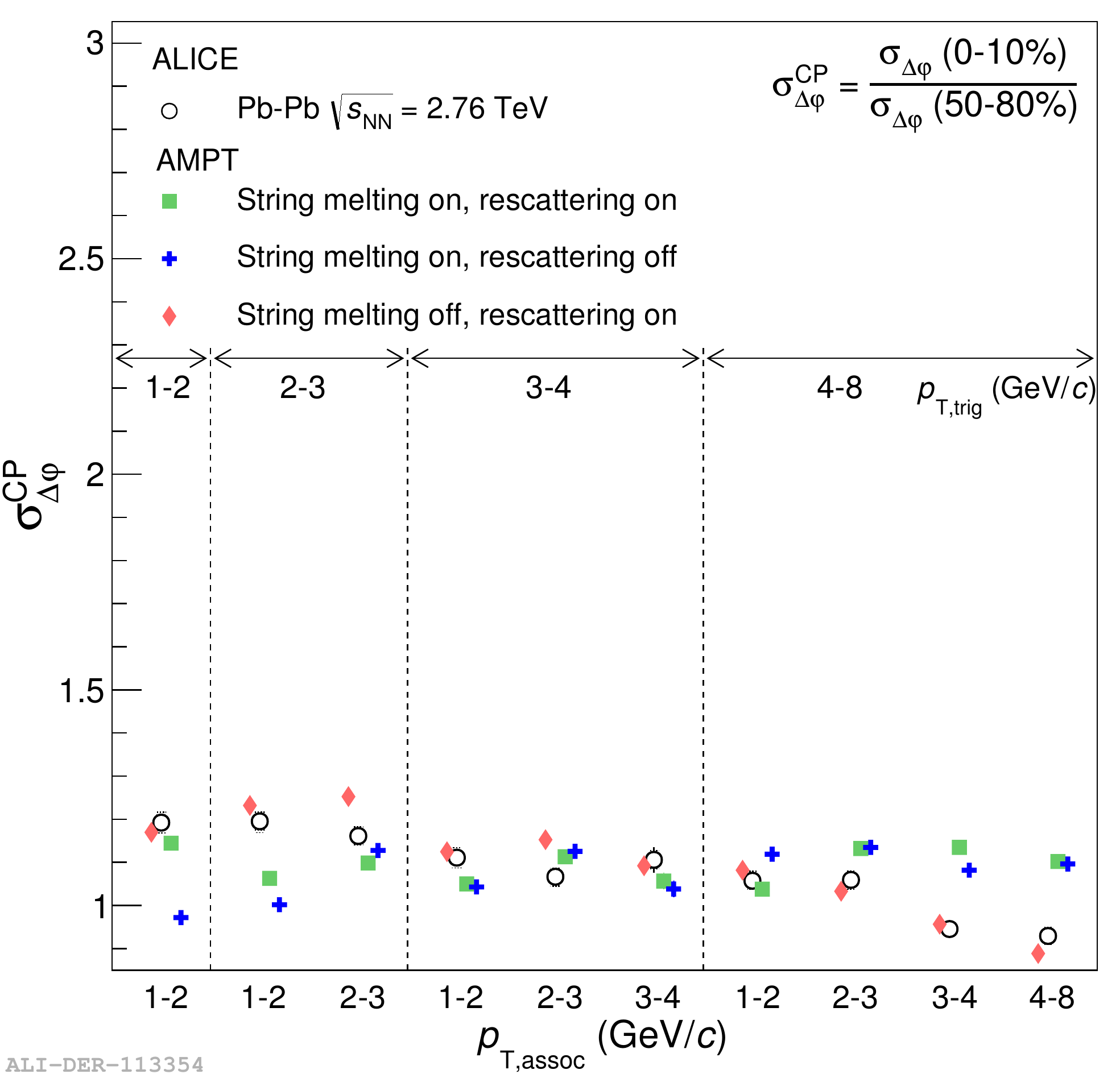}
    \end{overpic}\hfill
    \begin{overpic}[width=0.35\textwidth]{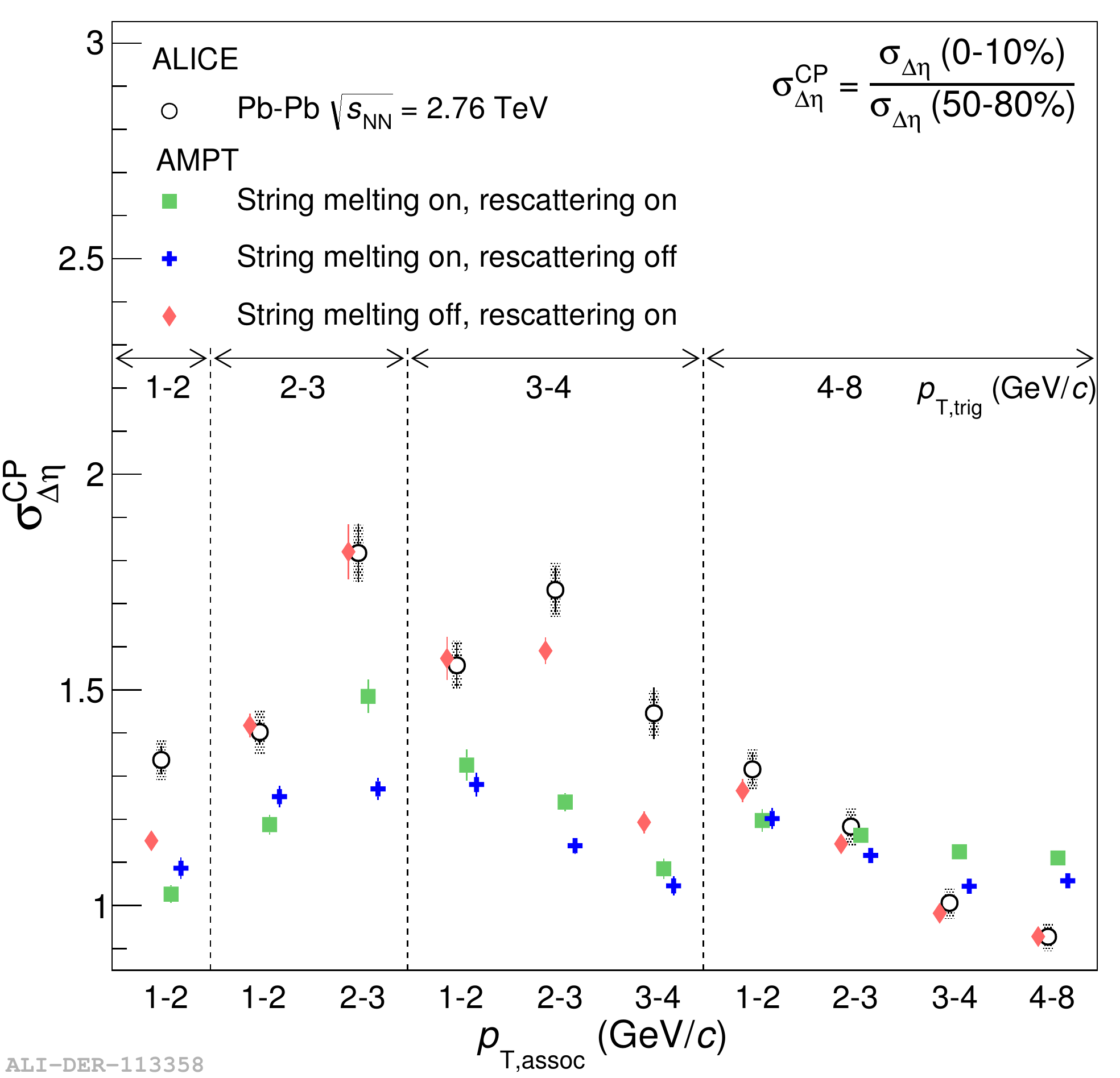}
    \end{overpic}\hfill}
  \caption{The left panel shows the missing yield in the depletion around $(\Delta\varphi,\Delta\eta) = (0,0)$ in two-particle number correlations as a function of centrality and compares it to simulations. The middle and right panels show the ratio of the width of two-particle number correlations in the most central (0--10\%) and in the most peripheral (50--80\%) bin from Pb--Pb collisions at $\sqrt{s_ {\rm NN}} = 2.76$~TeV compared to different settings of AMPT in the $\Delta\varphi$ (middle) and the $\Delta\eta$ (right) directions.}
  \label{fig:dip_CP}
\end{figure}

Number correlations at low $p_{\rm T}$ in the central cases exhibit a depletion around $(\Delta\varphi,\Delta\eta) = (0,0)$. This depletion has been quantified by calculating the missing fraction of the yield in the peak, and it is shown in the left panel of \cref{fig:dip_CP} for the centrality and $p_{\rm T}$ bins, where it is significantly different from 0. For the details of the calculation please see Refs.~\cite{PRL,PRC}.

\section{Comparison to Monte Carlo generators}
\label{sec:MC}
The results of particle number and transverse momentum correlations were compared to different settings of AMPT~\cite{AMPT} and to HIJING~\cite{HIJING}. The comparison of the width of the near-side peak in transverse momentum correlations can be seen in \cref{fig:width_pT}, where it is shown that no single model or setting can describe both directions and simultaneously the charge-dependent and the charge-independent case. 

The broadening and the depletion seen in two-particle number correlations are also compared to simulations. The broadening is described by the ratio of the width in the most central (0--10\%) and the most peripheral (50--80\%) bin. This is shown together with the different settings of AMPT in the middle and right panels of \cref{fig:dip_CP}. In the case of the two-particle number correlations, the broadening is well described by the AMPT setting with string melting turned off, but hadronic rescattering turned on; however, the absolute value of the width is not described better than 10\% by any of the settings (see Refs.~\cite{PRL,PRC} for details).

Simulations by AMPT also show a depletion if hadronic rescattering is turned on. The estimation of this depletion from AMPT is overlaid with the data in the left panel of \cref{fig:dip_CP}, and both settings shown describe the data within the uncertainties. 

\section{Conclusion}
Two-particle number and transverse momentum correlations from the ALICE collaboration were presented as a function of the collision centrality. The results were compared to HIJING and to different settings of AMPT. None of these models were able to reproduce all results, which shows that it is not enough to refine models using single-particle observable, but two-particle observables are needed as well.

This work has been supported by the Hungarian NKFIH/OTKA K 120660 grant.



\bibliographystyle{elsarticle-num}

\begin{thebibliography}{00}

 \bibitem{PRL} J. Adam et al. [ALICE collaboration], Phys. Rev. Lett. 119, 102301 (2017) 

 \bibitem{PRC} J. Adam et al. [ALICE collaboration], Phys. Rev. C 96, 034904 (2017)

 \bibitem{STAR} G. Agakishiev et al. [STAR Collaboration], Physics Letters B 704 (2011) 467-473

 \bibitem{AMPT} Z.-W. Lin, C. M. Ko, B.-A. Li, B. Zhang, and S. Pal, Phys. Rev. C72 (2005) 064901

 \bibitem{HIJING}  X.-N. Wang and M. Gyulassy, Phys. Rev. D44 (1991) 3501–3516.


 \end{thebibliography}



\end{document}